\documentclass[amsmath, a4paper, pra, twocolumn, floatfix, showpacs]{revtex4}
\usepackage{subfigure}
\usepackage{amssymb}
\usepackage{amsmath}
\usepackage{epsfig}
\usepackage{color}
\usepackage{graphics, graphicx}
\usepackage{psfrag}
\usepackage{mathcomp}

\begin{document}

\author{Michiel Snoek$^1$}
\author{Jun Liang Song$^2$}
\author{Fei Zhou$^2$}
\affiliation{$^1$Institute for Theoretical
Physics, Valckenierstraat 65, 1018 XE Amsterdam, the Netherlands \\
$^2$ Department of Physics and Astronomy,
The University of British Columbia, Vancouver, B. C., Canada V6T1Z1}



\title{Hyperfine Spin-Two ($F=2$) Atoms in Three-Dimensional Optical Lattices: \\
Phase Diagrams and Phase Transitions}

\begin{abstract}
We consider ultracold matter of spin-2 atoms in optical lattices. 
We derive an effective 
Hamiltonian for the studies of spin ordering in Mott states
and investigate hyperfine spin correlations. 
Particularly, we diagonalize the Hamiltonian in
an on-site Hilbert space taking into account spin-dependent interactions and exchange 
between different sites.
We obtain phase diagrams and quantum phase transitions between various magnetic
phases. 
\end{abstract}

\maketitle

\section{Introduction}

Optically trapped ultracold atoms possess various hyperfine spin degrees of freedom.
By trapping spinor atoms in optical lattices, one can explore the physics of 
hyperfine spin correlated ultracold atomic matter which generally has 
fantastically rich magnetic properties. It had been pointed out a while ago that
cold atoms in optical lattices can be used to simulate correlated physics related to 
the Bose-Hubbard model \cite{Jaksch98} and 
the superfluid-Mott insulator transition; this transition, tuned by optical lattice potential depth, 
has been observed in recent experiments \cite{Greiner02}.
For cold atoms with hyperfine spins, additional magnetic transitions in 
Mott states are 
possible. For instance, spin-one atoms can have either ferromagnetic interactions, 
such as for $^{87}$Rb, or antiferromagnetic interactions, 
as for $^{23}$Na \cite{Ho98,Ohmi98,Stenger98}. 
As shown before for sodium atoms, 
a first order phase transition between 
spin-ordered (nematic) and spin-disordered (spin-singlet) 
ground state occurs \cite{Demler02, Zhou03, Imambekov03,Zhou03b, Snoek04} in the Mott insulating state. 
Magnetic fields or magnetization can further induce
spontaneous nematic ordering \cite{Imambekov04,Zhou04}. 

Correlated spin-2 atoms have also been studied recently
and were suggested to possess even richer phases. 
Apart from nematic and ferromagnetic phases, a cyclic phase is also 
proposed \cite{Ciobanu00, Ueda02, Barnett06}.
Fascinating fractionalized non-abelian vortex structures have been predicted \cite{Semenoff06, Pogosov05}. 
For nematic condensates, the accidental degeneracy between nematic 
states with different symmetries (i.e. uni-axial versus bi-axial) has been shown to be lifted by zero point 
energies of spin-wave excitations 
which effectively can be attributed to the mechanism of order-from-disorder \cite{Song07, Turner07}. 
In lattices, interactions between spin-two atoms further give rise to new kinds of spin-ordered to 
spin-disordered transitions \cite{Zhou06}. A new class of
quantum coherent dynamics induced by quantum fluctuations of spin waves and
tuned by the optical lattice potential depth was also investigated recently \cite{Song07b}. 
The distinct magnetic correlations in
Mott insulating states were not taken into account in early studies of
Mott-superfluid transitions of spin-2 atoms \cite{Hou03, Jin04}.

Experimentally, atoms on an $F=2$ manifold are usually less stable than those on an $F=1$ manifold 
when spin-one multiplets 
are lower in energy and spin-flip scattering processes lead to quick relaxation of $F=2$ atoms.
This is particularly problematic for $F=2$ multiplets of $^{23}$Na where spin-flip scattering 
is quite strong. But for $^{87}$Rb, spin-flip scattering is relatively weak. 
This isotope is therefore a more likely candidate for observance of the physics of correlated spin-two atoms. 
Early measurements \cite{Roberts98} and theoretical calculations of scattering lengths \cite{Klausen01} suggest 
that spin-2 
$^{87}$Rb atoms 
have a nematic ground state.
Coherent spin dynamics in condensates of spin-one or spin-two cold atoms as well as
few-body controlled collisions have already 
been studied in experiments
\cite{Schmaljohann04, Chang04, Widera05, Widera06}, 
although direct evidence of spin correlated ultra cold matter in optical lattices is still absent.
Investigation of cold atoms with high spins
in optical lattices will lead to better understanding of 
fundamental principles of quantum magnetism; in addition, 
it might also lead to potential applications towards quantum 
information
storing and processing \cite{Kitaev03, Raussendorf01}.

In this article, we present detailed analysis of quantum states of 
spin-2 atoms in optical lattices. 
This subject was also addressed in a previous work
where the authors minimized the mean-field energies of 
maximally ordered states with respect to a tensor order 
parameter \cite{Zhou06};
those trial wavefunctions approximate the ground states quite well in the limit of large 
exchange coupling  but deviations from those states can be substantial in the intermediate
coupling regime. To address all possible ordered phases, in this article we
extend our analysis to all possible mean-field states and
carry out a systematic calculation to further determine the phase boundaries and the order of the  
phase transitions. We also discuss quadratic Zeeman effects. 

The organization of this paper is as follows.
In Section II we define the 
system and operator-algebra.  In 
Section III we consider the limit of zero hopping 
and derive exact phase diagrams for arbitrary numbers of atoms per 
lattice site. 
In Section IV we describe 
the self-consistent mean-field technique to deal with nonzero exchange coupling. In Section V we do calculations for nonzero exchange between the sites using this mean-field method for two, three and four particles per site. 
 We conclude our studies in Section VI.

\section{Description of the system}
In this section we introduce the theoretical framework to deal with cold gases of $F=2$ atoms. We describe the algebra of the number and spin operators and derive the Hamiltonian.

\subsection{Algebra}
As a starting point we take the usual creation and annihilation operators for
$F=2, m=-2,\ldots, 2$ particles:
\begin{equation}
\hat \psi_m, \quad \hat \psi_m^\dagger, \quad \lbrack \hat \psi_m, \hat \psi_{m'}^\dagger \rbrack =
\delta_{m m'}.
\end{equation}
To work conveniently with the hopping term in the Hamiltonian we now introduce another basis. Using the spherical harmonics $Y_{2m} (\theta, \phi)$, the following operators are
constructed:
\begin{eqnarray}
\hat \psi_{xx} &=& \frac{1}{\sqrt{2}}(\hat \psi_{-2} + \hat \psi_2) - \frac{1}{\sqrt{3}}
\hat \psi_0, \\
\hat \psi_{yy} &=& -\frac{1}{\sqrt{2}}(\hat \psi_{-2} + \hat \psi_2) - \frac{1}{\sqrt{3}}
\hat \psi_0, \\
\hat \psi_{zz} &=&  \frac{2}{\sqrt{3}} \hat \psi_0, \\
\hat \psi_{xy} &=& \frac{i}{\sqrt{2}}(\hat \psi_{-2} - \hat \psi_2)  \\
\hat \psi_{xz} &=& \frac{1}{\sqrt{2}}(\hat \psi_{-1} - \hat \psi_1)  \\
\hat \psi_{yz} &=& -\frac{i}{\sqrt{2}}(\hat \psi_{-1} + \hat \psi_1), 
\end{eqnarray}
and the creation operators in the same way. These operators have the following
properties \cite{Zhou06}:
\begin{eqnarray}
\lbrack \hat \psi_{\alpha \beta}, \hat \psi_{\alpha' \beta'}^\dagger \rbrack &=& 
\delta_{\alpha \alpha'} \delta_{\beta \beta'} + 
\delta_{\alpha \beta'} \delta_{\beta \alpha'} -
\frac{2}{3} \delta_{\alpha \beta} \delta_{\alpha' \beta'} \\
{\rm Tr}[\hat \psi] &=& \sum_{\alpha} \hat \psi_{\alpha \alpha} = 0.
\end{eqnarray}
This last property puts a constraint on the constructions of linear 
operators. For operators 
\begin{equation}
{\rm Tr} \lbrack \Delta \hat \psi \rbrack = \sum_{\alpha, \beta} \Delta_{\alpha
\beta} \hat \psi_{\beta \alpha},
\end{equation}
the tensor $\Delta$ can be always reduced to a traceless one, i.e.,
\begin{equation}
{\rm Tr} [ \Delta] = \sum_{\alpha} \Delta_{\alpha \alpha} = 0. 
\end{equation}
This constraint is needed, because when introducing this new basis we 
have enlarged the Hilbert space by constructing six operators out of five. This constraint brings the size of the physical Hilbert space back to the original one.

The density operator in terms of the new operators can be derived as:
\begin{equation}
\hat \rho = \sum_m \hat \psi_m^\dagger \hat \psi_m = \frac{1}{2} {\rm Tr}[ \hat \psi^\dagger
\hat \psi ] = 
\frac{1}{2} \sum_{\alpha, \beta} \hat \psi^\dagger_{\alpha \beta} \hat \psi_{\beta
\alpha}.
\end{equation}
The factor $\frac{1}{2}$ 
appears here because the trace involves a double 
sum over the operator $\hat \psi_{\alpha\beta}$. 
This same factor will appear later when 
deriving the hopping term in the Hamiltonian.

The spin operator is straightforwardly derived as:
\begin{equation}
\hat F_\alpha  = - i \epsilon_{\alpha \beta \gamma} \hat \psi_{\beta \eta}^\dagger
\hat \psi_{\eta \gamma}.
\end{equation}
It has the following properties:
\begin{eqnarray}
\lbrack \hat F_\alpha, \hat F_\beta \rbrack &=& i \epsilon_{\alpha \beta \gamma} \hat F_\gamma, \\
\lbrack \hat F_\alpha, \hat \rho \rbrack &=& 0 \\
\lbrack \hat F_\alpha, {\rm Tr}[(\hat \psi^\dagger)^n ] \rbrack &=& 0.
\end{eqnarray}

The total spin operator is then given by:
\begin{eqnarray}
\hat F^2 &=& \hat F_\alpha \hat F_\alpha \\ &=& 
\hat \psi_{\beta \eta}^\dagger \hat \psi_{\eta \gamma}
\hat \psi_{\gamma \xi}^\dagger \hat \psi_{\xi \beta} -
\hat \psi_{\beta \eta}^\dagger \hat \psi_{\eta \gamma}
\hat \psi_{\beta \xi}^\dagger \hat \psi_{\xi \gamma} \\
&=&
\hat \psi_{\beta \eta}^\dagger \hat \psi_{\gamma \xi}^\dagger  
\hat \psi_{\eta \gamma} \hat \psi_{\xi \beta} -
\hat \psi_{\beta \eta}^\dagger \hat \psi_{\beta \xi}^\dagger 
\hat \psi_{\eta \gamma} \hat \psi_{\xi \gamma} 
+ 6 \hat \rho.
\end{eqnarray}

We also introduce the dimer creation operator as:
\begin{equation}
\hat{\mathcal{D}^\dagger} = \frac{1}{\sqrt{40}} {\rm Tr} [ (\hat \psi^\dagger)^2 ],
\end{equation}
which has the following properties:
\begin{equation}
\lbrack \hat{\mathcal{D}}, \hat{\mathcal{D}}^\dagger \rbrack = 1 + \frac{2}{5} \hat \rho.
\end{equation}
This operator creates two particles which together form a spin singlet. In the same way we can construct an operator which creates three particles which together form a singlet. This is called the trimer operator and defined as
\begin{equation}
\hat{\mathcal{T}}^\dagger = \frac{1}{\sqrt{140}} {\rm Tr} [ (\hat \psi^\dagger)^3 ].
\end{equation}

Finally we introduce the nematic operator as:
\begin{equation}
\hat Q_{\alpha \beta} = \hat \psi_{\alpha \eta}^\dagger \hat \psi_{\eta \beta}  - \frac{1}{3} \delta_{\alpha \beta} {\rm Tr} [ \hat \psi^\dagger \hat \psi] 
=  \hat \psi_{\alpha \eta}^\dagger \hat \psi_{\eta \beta} - \frac{2}{3} \delta_{\alpha \beta} \hat \rho.
\end{equation}
The non-vanishing eigenvalues of this operator indicate the presence of 
nematic order 
\cite{Snoek04}.

\subsection{Hamiltonian}
We consider  $F=2$ atoms in an optical lattice. The laser wavelength is $\lambda$. This results in a potential  $V ({\bf r}) = V_0 (\sin^2 (2 \pi x/\lambda) + \sin^2 (2 \pi y/\lambda) + \sin^2 (2 \pi z /\lambda) )$. We assume that the optical lattice potential is deep enough such that the lowest band approximation and the tight binding approximation are applicable. The Hamiltonian is 
then given as \cite{Zhou06}:
\begin{eqnarray}
\hat{\mathcal{H}} &=& 
\frac{a_L}{2} \sum_i (\hat \rho_i^2 - \hat \rho_i) + 
\frac{b_L}{2} \sum_i (\hat F_i^2 - 6 \hat \rho_i) \nonumber \\ && +
5 c_L \sum_i \hat{\mathcal{D}}_i^\dagger \hat{\mathcal{D}}_i
- t \sum_{\langle i j \rangle} {\rm Tr} [ \hat \psi_i^\dagger \hat \psi_j ], 
\label{ham0}
\end{eqnarray}
where $i$ is the site index, $\langle i j \rangle$ means that the sum is over
neighboring sites, $t$ is the hopping parameter and the constants $a_L$, $b_L$
and $c_L$ can be expressed, in terms of atomic mass $M$, on-site ground state
wavefunction $\hat \psi_0({\bf x})$ and scattering lengths $a_F$ in the total
hyperfine spin $F=0,2,4$ channels, as:
\begin{eqnarray}
a_L &=& \frac{4 \pi \hbar^2(4 a_2+3 a_4)}{7 M} \int d^3 {\bf x} |\hat \psi_0 ({\bf
x})|^4 \\
b_L &=& \frac{4 \pi \hbar^2(a_4 - a_2)}{7 M} \int d^3 {\bf x} |\hat \psi_0 ({\bf
x})|^4 \\
c_L &=& \frac{4 \pi \hbar^2(7 a_0 - 10 a_2 +3 a_4)}{35 M} \int d^3 {\bf x}
|\hat \psi_0 ({\bf x})|^4.  
\end{eqnarray}
The hopping amplitude $t$ is given by the overlap integral
\begin{equation}
t = - \frac{1}{2} \int d^3 {\bf x}  \hat \psi_0 ({\bf x}) \left[ - \frac{ \hbar^2 \nabla^2}{2 m} + V({\bf x})
\right]\hat \psi({\bf x}+ \frac{\lambda}{2}{\bf e}_i),
\end{equation}
where ${\bf e}_i$ are the unit-vectors in $x$, $y$ and $z$ direction. Note the additional factor $\frac{1}{2}$ appearing here. This factor needs to be inserted, because the trace in the hopping term in Eq. (\ref{ham0}) involves a double sum over the indices of the creation and annihilation operators $\hat \psi_{\alpha \beta}^{(\dagger)}$.

\subsection{Mott Hamiltonian for $\rho>1$} 
We now assume that the system is in a Mott state with $\rho$ particles per site. 
When the number of particles on a lattice site is larger than one (i.e. $\rho>1$), we assume
that the spin splitting in the virtual hopping process can be ignored.  This is justified because $a_L \gg b_L, c_L$, 
such that the density-density interaction dominates. This leads to an effective Mott
Hamiltonian 

\begin{eqnarray} \label{ham}
\hat{\mathcal{H}}_{\rm Mott} &=& 
\frac{b_L}{2} \sum_i (\hat F_i^2 - 6 \hat \rho_i) +
5 c_L \sum_i \hat{\mathcal{D}}_i^\dagger \hat{\mathcal{D}}_i 
\\ &&
- J_{\rm ex} \sum_{\langle i j \rangle} \left( \hat \psi_{i, \alpha \beta}^\dagger \hat \psi_{j,
\beta \alpha}
\hat \psi_{j, \alpha' \beta'} \hat \psi_{i, \beta' \alpha'}^\dagger+ {\rm h.c.} \right), \nonumber
\end{eqnarray}
where $J_{\rm ex} = t^2/a_L$ is the exchange coupling.

In analogy with the spin $F=1$ case, we now introduce the 'traceless' operator
\begin{eqnarray}
&& \hat Q_{i; \alpha, \beta, \alpha' \beta'}^\dagger = 
 \hat \psi_{i, \alpha \beta}^\dagger \hat \psi_{i, \alpha' \beta'} \\&& -
 \frac{1}{10} {\rm Tr}[\hat \psi_i^\dagger \hat \psi_i] 
 \left( 
 \delta_{\alpha \alpha'} \delta_{\beta \beta'} +
 \delta_{\alpha \beta'} \delta_{\beta \alpha'}  - 
 \frac{2}{3} \delta_{\alpha \beta} \delta_{\alpha' \beta'} \right). \nonumber
\end{eqnarray}
Using the definition of $\hat \rho$, we can rewrite:
\begin{eqnarray}
&& \hat Q_{i; \alpha, \beta, \alpha' \beta'}^\dagger = \\ &&
 \hat \psi_{i, \alpha \beta}^\dagger \hat \psi_{i, \alpha' \beta'} -
 \frac{1}{5} \hat \rho_i 
 \left( 
 \delta_{\alpha \alpha'} \delta_{\beta \beta'} +
 \delta_{\alpha \beta'} \delta_{\beta \alpha'}  - 
 \frac{2}{3} \delta_{\alpha \beta} \delta_{\alpha' \beta'} \right). \nonumber
\end{eqnarray}
This operator is 'traceless' because
\begin{equation}
\sum_{\alpha, \beta} \hat Q_{i; \alpha \beta, \beta, \alpha}^\dagger = 0.
\end{equation}
It has the property that is symmetric under interchange of $\alpha$ and $\beta$
and $\alpha'$ and $\beta'$ and that
\begin{equation}
\hat Q_{i; \alpha \beta, \alpha' \beta'} = \hat Q_{i; \alpha' \beta', \alpha
\beta}^\dagger.
\end{equation}
In terms of this operator the exchange term (due to virtual hopping processes) in the Hamiltonian can
be rewritten as (up to terms which contain the local density and in the Mott
state therefore only give rise to an energy shift):
\begin{eqnarray} 
\hat{\mathcal{H}}_{\rm ex} &=& -J_{\rm ex} \sum_{\langle i j \rangle} \left( \hat Q_{i; \alpha \beta, \alpha'
\beta'}^\dagger
\hat Q_{j; \alpha \beta, \alpha' \beta'} + {\rm h.c.} \right). 
\end{eqnarray}

\subsection{Magnetic Fields: Quadratic Zeeman Effect}
The presence of a magnetic field leads to linear and quadratic Zeeman effects. The linear Zeeman effect leads to an additional term in the Hamiltonian
\begin{equation}
\hat{\mathcal{H}}_{\rm lin. Z} = - q_{\rm lin} \sum_i {\bf B} \cdot \hat {\bf F}_i.
\end{equation}
For concreteness we take the magnetic field in the $z$-direction: ${\bf B}  = B e_z$ and  $\hat{\mathcal{H}}_{\rm lin. Z} = - q_{\rm lin} B \sum_i \hat F_{i,z}$.
However, the total Hamiltonian commutes with $\hat F_{i,z}$, so that once the system is prepared, 
the expectation value $\langle \hat F_{i,z} \rangle$ will remain the same. In experiments,  atoms are usually initially prepared in 
the $(2,0)$-state. This means that in the experimental situation the linear Zeeman effect is irrelevant. 
Relevant is the quadratic Zeeman effect. It is important to note that the quadratic Zeeman effect gives an energy shift to the individual particles, depending on their spin-state. Writing $\hat n_{i,m} = \hat \psi_{i,m}^\dagger \hat \psi_{i,m}$, $m= -2, \ldots, 2$, the Hamiltonian describing the quadratic Zeeman effect is therefore given by:
\begin{equation}
\hat{\mathcal{H}}_{\rm quad. Z} = q_{\rm quad} \sum_i \left(\hat  n_{i,1} + \hat n_{i,-1}  + 4(\hat n_{i,2} + \hat n_{i,-2}) \right).
\end{equation}
Observing now that 
\begin{equation}
\hat Q_{i,zz} = - \frac{1}{3} ( \hat n_{i,1} + \hat n_{i,-1})  -  \frac{4}{3}(\hat n_{i,2} + \hat n_{i,-2}) + \frac{2}{3} \hat \rho_i
\end{equation}
we see that we can write:
\begin{equation}
\hat{\mathcal{H}}_{\rm quad. Z} = - 3 q_{\rm quad} \sum_{i}\hat Q_{i,zz},
\end{equation}
where we leave out the term involving $\hat \rho_i$ because it only gives a constant contribution.

Like in the case of spin-1 bosons \cite{Zhou04} we see that this term does not commute with $\hat F^2$. Therefore the spin-singlet states are unstable with respect to this perturbation and nematic order is induced for infinitesimally small coupling.

\section{On-site spectrum}
When the tunneling is zero, the sites are decoupled. In this case the full spectrum can
be derived for arbitrary (integer) numbers of particles per site 
\cite{Ueda02, Hou03}.
The local operators $\hat \rho$, $\hat F_\alpha$, $\hat{\mathcal{D}}$ and
$\hat{\mathcal{D}}^\dagger$ obey the following commutation relations
\begin{eqnarray*}
\lbrack \hat  F_\alpha, \hat \rho \rbrack &=& 0 \\
\lbrack \hat F_\alpha, \hat{\mathcal{D}} \rbrack &=& 0 \\
\lbrack \hat F_\alpha, \hat{\mathcal{D}}^\dagger \rbrack &=& 0 \\
\lbrack \hat F_\alpha, \hat F_\beta \rbrack &=& i \epsilon_{\alpha \beta \gamma} \hat F_\gamma, \\
\lbrack \hat{\mathcal{D}}, \hat \rho \rbrack &=& 2 \hat{\mathcal{D}} \\
\lbrack \hat{\mathcal{D}}^\dagger, \hat \rho \rbrack &=& - 2 \hat{\mathcal{D}}^\dagger \\
\lbrack \hat{\mathcal{D}}, \hat{\mathcal{D}}^\dagger \rbrack &=& 1 + \frac{2}{5} \hat \rho 
\end{eqnarray*}
The spin-operators commute with all other local operators and form a
SU(2)-algebra. The density and dimer operators together form a
$SU(1,1)$-algebra \cite{Ueda02}. This can be seen by defining 
\begin{eqnarray}
\hat{\mathcal{D}}^- &=& \sqrt{\frac{5}{2}} \hat{\mathcal{D}} \\
\hat{\mathcal{D}}^+ &=& \sqrt{\frac{5}{2}} \hat{\mathcal{D}}^\dagger\\
\hat{\mathcal{D}}^z &=& \frac{\hat \rho}{2} + \frac{5}{4}.
\end{eqnarray}
Those operators obey the algebra:
\begin{eqnarray}
\lbrack \hat{\mathcal{D}}^z, \hat{\mathcal{D}} ^\pm \rbrack &=& \pm \hat{\mathcal{D}} ^\pm \\
\lbrack \hat{\mathcal{D}}^+, \hat{\mathcal{D}}^- \rbrack &=& - 2 \hat{\mathcal{D}}^z
\end{eqnarray}
In analogy with the spin-algebra we now define the Casimir operator
$\hat D^2$ as
\begin{eqnarray}
\hat D^2 &=& - \frac{1}{2} \left( \hat{\mathcal{D}}^- \hat{\mathcal{D}}^+ +
\hat{\mathcal{D}}^+ \hat{\mathcal{D}}^- \right) + \hat{\mathcal{D}}^z \hat{\mathcal{D}}^z \\
&=&  - \frac{5}{2} \hat{\mathcal{D}}^\dagger \hat{\mathcal{D}}  - \left( \frac{5}{4} +
\frac{\hat \rho}{2} \right) +  \left( \frac{5}{4} + \frac{\hat \rho}{2}
\right)^2.
\end{eqnarray}
This operator commutes with $\hat{\mathcal{D}}^\pm$ and $\hat{\mathcal D}^z$ and therefore
also with $\hat{\mathcal D}^\dagger$, $\hat{\mathcal{D}}$ and $\hat \rho$.
Now we consider a state $|\psi_{\rm min} \rangle$ with $\rho_{\rm min}$ atoms (i.e. $ \hat \rho | \psi_{\rm min} \rangle = \rho_{\rm min}  | \psi_{\rm min} \rangle$ and
$\hat{\mathcal{D}} |\psi_0 \rangle = 0$. 
The operator $\hat{\mathcal{D}}$ destroys two atoms that form a singlet pair. 
That means that the above introduced state 
$|\psi_{\rm min} \rangle $ 
contains no singlet pairs and $\rho_{\rm min}$ unpaired atoms. 
Applying $\hat D^2$ to this state we get
$ \hat D^2
|\psi_{\rm min} \rangle = \lambda_D (\lambda_D + 1) |\psi_{\rm min} \rangle$
with $\lambda_D = \frac{ 1 + 2 \rho_{\rm min}}{4}$. 
Applying now $\hat{\mathcal{D}}^\dagger$ to $|\psi_{\rm min}\rangle$ adds singlet 
pairs 
but keeps the number of unpaired atoms constant. 
Since $\hat D^2$
commutes with $\hat{\mathcal{D}}^\dagger$, 
all the states
$\left(\hat{\mathcal{D}}^\dagger\right)^n |\psi_{\rm min} \rangle$ with number of
atoms $\rho=\rho_{\rm min} + 2 n_p$ 
(i.e. $\rho_{\rm min}$ unpaired atoms and $n_p$ pairs)
have the same quantum number
$\lambda_D = \frac{1 + 2 \rho_{\rm min}}{4}$ of the Casimir operator. 

So we find a relation between the particle number and the quantum number of the
Casimir operator as $\rho = \frac{4 \lambda_D - 1}{2} + 2 n_p$ 
which means that for $\rho$ atoms, the possible eigenvalues are 
$\lambda_D = \frac{2 \rho + 1}{4} - n_p$, $n_p$ being the number of pairs.  
We therefore replace the quantum number $\lambda_D$ by the number 
of unpaired atoms $d = \rho_{\rm min} = \frac{ 4 \lambda_D - 1}{2}$.

\begin{figure}
\includegraphics[scale=.7]{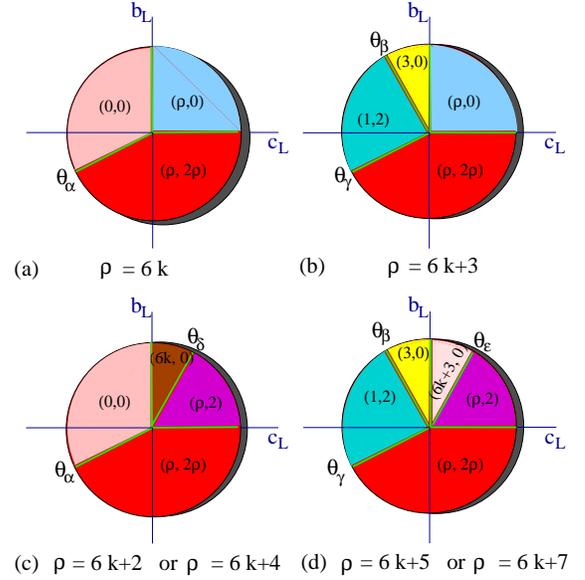}
\caption{(Color online) Phase diagrams in the case of isolated lattice sites for  
various particles numbers $\rho$. The different phases are indicated by the quantum numbers $d$ and $F$:  $(d,F)$.
The angles $\theta_\alpha, \ldots, \theta_\xi$ between the phases depend on the particle number and are defined in the text.}
\label{phasediagj=0}

\end{figure}

We can then express the on-site energy in terms of the four quantum numbers $\rho$, $d$ and $F$. After some straightforward algebra this yields:
\begin{eqnarray}
E(\rho, d, F) &=& \frac{a_L}{2} \rho( \rho-1) + \frac{b_L}{2} (F(F+1) - 6 \rho) \nonumber \\ && + \frac{c_L}{8} \left( (2 \rho + 3)^2 - ( 2 d + 3)^2 \right). 
\end{eqnarray}
In order to find the ground state, we have to take care of the bosonic symmetry. The requirement that the bosonic wave function is symmetric implies that some combinations of quantum numbers are forbidden. 
In general for a number of atoms $\rho$ we have $\rho = d + 2 n_p$ and 
$F = 0, \ldots, 2 d$, because the paired atoms don't contribute to the spin. 
However, if $d = 3 k $ the values $F=1,2,5,2d-1$ 
are forbidden because of symmetry and if 
$d = 3 k \pm 1$ the values $F= 0,1,3,2d-1$ are forbidden. 
By minimizing the energy under those conditions the ground 
states can be identified. We label each state by two quantum
numbers as $(d,F)$.
This yields the phase diagrams as shown in Fig. \ref{phasediagj=0}.

The phases are separated by critical angles $\tan \theta = \frac{b_L}{c_L}$, which are given by:
\begin{eqnarray}
\tan \theta_{\alpha}  &=& \frac{\rho+3}{4 \rho +2 } \\
\tan \theta_{\beta}  &=& - \frac{7}{3} \\
\tan \theta_{\gamma}  &=& \frac{\rho+4}{4 \rho +6 } \\
\tan \theta_{\delta}  &=& 4k+\frac{5}{3} \mbox{ if } \rho = 6k+2 \\
\tan \theta_{\delta}  &=& 8k+\frac{14}{3} \mbox{ if } \rho = 6k+4 \\
\tan \theta_{\epsilon}  &=& 4k + \frac{11}{3}  \mbox{ if } \rho = 6k+5 \\
\tan \theta_{\epsilon}  &=& 8k + \frac{26}{3}  \mbox{ if } \rho = 6k+7 
\end{eqnarray}

The states appearing in this limit are

\begin{itemize}
\item{Ferromagnetic:} $d = \rho$, $F = 2 d = 2 \rho$ 
\item{Trimer:} $d = \rho = 3 k$, $F = 0$
\item{Cyclic:} $d = \rho = 3 k \pm 1$, $F = 2$
\item{Dimer:} $d = 0$, $F=0$
\item{Nematic:} $d = 1$, $F=2$.
\end{itemize}
It is worth remarking that strictly speaking for individual lattice site, there is 
no long range order 
and symmetry-breaking states do not exist in this limit.
However, infinitesimal hopping could couple 
the directors of the broken symmetries in some cases
and establish long range order; 
the notations of {\em nematic} and {\em cyclic} introduced above refer to states 
which will have the respective long range order if infinitesimal hopping is allowed
and are only truly meaningful when nonzero hopping is taken into account.

In the numerical scheme pursued in the next sections, we distinguish the phases by the following order parameters:
\begin{itemize}
\item{Ferromagnetic:} $\langle \hat{\mathcal{D}}^\dagger \hat{\mathcal{D}} \rangle= 0$, $\langle \hat F^2 \rangle = 2 \rho (2 \rho + 1)$.
\item{Trimer:} $\langle \hat F^2 \rangle = 0$, $\langle \hat{\mathcal{D}}^\dagger \hat{\mathcal{D}} \rangle = 0$, $\langle \hat Q_{\alpha \beta} \rangle = 0$.
\item{Cyclic:} $\langle \hat{\mathcal{D}}^\dagger \hat{\mathcal{D}} \rangle = 0$, $0<\langle \hat F^2 \rangle < 2 \rho (2 \rho + 1)$, $\langle \hat Q_{\alpha \beta} \rangle = 0$.
\item{Dimer:}  $\langle \hat F^2 \rangle = 0$, $\langle \hat{\mathcal{D}}^\dagger \hat{\mathcal{D}} \rangle = \frac{\rho(\rho+3)}{10}$, $\langle \hat Q_{\alpha \beta} \rangle = 0$.
\item{Nematic:} $\langle \hat F_\alpha \rangle = 0$, $\langle \hat Q_{\alpha \beta} \rangle \neq 0$, $\langle \hat{\mathcal{D}}^\dagger \hat{\mathcal{D}} \rangle > 0$.
\end{itemize}

\section{Nonzero tunneling}
We now turn to the case of nonzero tunneling between neighboring lattice sites. In this case there is a competition between states with broken symmetries or long range order and states without broken symmetries. To deal with this situation we make the Ansatz that the total many-body wave function is a product wave function over the lattice sites:
\[
| \Psi_{\rm tot} \rangle = \prod_i |  \Psi_i \rangle_i.
\]
We moreover assume that the spatial symmetry is unbroken, such that the wavefunctions are identical on each lattice site. We thereby exclude antiferromagnetically ordered states, but they turn out to have higher energy than the states with unstaggered long range order. In the numerical scheme they would moreover be identified by oscillating solutions.
Following this procedure the Hamiltonian in Eq. (\ref{ham}) turns into a local Hamiltonian, which is coupled in mean-field to the neighboring lattice sites:
\begin{eqnarray} \label{hammf}
&& \hat{\mathcal{H}}_{\rm MF} = 
\frac{b_L}{2} \sum_i (\hat F_i^2 - 6 \hat \rho_i) +
5 c_L \sum_i \mathcal{D}_i^\dagger \mathcal{D}_i 
\nonumber
\\ &&
- J_{\rm ex} \sum_{\langle i j \rangle} \left( 
\hat Q_{i; \alpha \beta, \alpha' \beta'}^\dagger Q_{j; \alpha \beta, \alpha' \beta'} + 
Q_{i; \alpha \beta, \alpha' \beta'}^* \hat Q_{j; \alpha \beta, \alpha' \beta'} \right. \nonumber \\ && \left.
\hspace{1cm} - Q_{i; \alpha \beta, \alpha' \beta'}^* Q_{j; \alpha \beta, \alpha' \beta'} + {\rm h.c.} \right),
\end{eqnarray}
where $Q_{j; \alpha \beta, \alpha' \beta'} = \langle \hat Q_{j; \alpha \beta, \alpha' \beta'} \rangle$.  The term $ J_{\rm ex} Q_{i; \alpha \beta, \alpha' \beta'}^* Q_{j; \alpha \beta, \alpha' \beta'}$ is a constant term in the Hamiltonian. However, this term is important for comparing energies of the different states, to be able to identify the ground state solution in the case of multiple stable solutions. 

Since this is now only a local problem we drop the site index and get (also dropping the constant terms):

\begin{eqnarray} \label{hamloc}
&& \hspace{-1cm} \hat{\mathcal{H}}_{\rm local} = 
\frac{b_L}{2} (\hat F^2 - 6 \hat \rho) +
5 c_L \hat{\mathcal{D}}^\dagger \hat{\mathcal{D}} 
\nonumber
\\ && \hspace{-1cm}
- z J_{\rm ex} \left( 
\hat Q_{\alpha \beta, \alpha' \beta'}^\dagger Q_{\alpha \beta, \alpha' \beta'} + 
 \hat Q_{\alpha \beta, \alpha' \beta'} Q_{\alpha \beta, \alpha' \beta'}^* \right).
\end{eqnarray}
Here we have introduced the lattice coordination number $z$, which is equal to $z=6$ for the three-dimensional cubic lattice.

To take into account the full on-site Hilbert space, we use another basis.
Namely, we define five symmetric, traceless tensors $\Delta_{\mu}$, which are
orthonormal in the sense that
\begin{equation}
{\rm Tr}[\Delta_{\mu}^* \Delta_{\nu} ] = \delta_{\mu \nu}/2.
\end{equation}
An explicit example of these are given by:
\begin{eqnarray}
\Delta_1 &=& \frac{1}{2\sqrt{3}} \left( \begin{array}{ccc} 1 & 0 & 0 \\ 0 & 1 &
0 \\ 0 & 0 & -2 \end{array} \right), 
\quad
 \Delta_2 = \frac{1}{2} \left( \begin{array}{ccc} 1 & 0 & 0 \\ 0 & - 1 & 0 \\ 0
& 0 & 0 \end{array} \right), 
 \\
 \Delta_3 &=& \frac{1}{2} \left( \begin{array}{ccc} 0 & 1 & 0 \\ 1 &  0 & 0 \\ 0 &
0 & 0 \end{array} \right), \quad
  \Delta_4 = \frac{1}{2} \left( \begin{array}{ccc} 0 & 0 & 1 \\ 0 &  0 & 0 \\
1 & 0 & 0 \end{array} \right)
  \\
   \Delta_5 &=& \frac{1}{2} \left( \begin{array}{ccc} 0 & 0 & 0 \\ 0 &  0 & 1 \\ 0
& 1 & 0 \end{array} \right) 
\end{eqnarray}
This choice is arbitrary, but has the advantage that we can work with purely real matrices. In terms of the original spin-operators we have:
\begin{eqnarray*}
{\rm Tr} [ \Delta_1 \hat \psi ] &=& - \hat \psi_0 \\
{\rm Tr} [ \Delta_2 \hat \psi ] &=& \frac{1}{\sqrt{2}} \left( \hat \psi_{-2} + \hat \psi_{2} \right) \\
{\rm Tr} [ \Delta_3 \hat \psi ] &=& \frac{i}{\sqrt{2}} \left( \hat \psi_{-2} - \hat \psi_{2} \right) \\
{\rm Tr} [ \Delta_4 \hat \psi ] &=& \frac{1}{\sqrt{2}} \left( \hat \psi_{-1} - \hat \psi_{1} \right) \\
{\rm Tr} [ \Delta_5 \hat \psi ] &=& -\frac{i}{\sqrt{2}} \left( \hat \psi_{-1} + \hat \psi_{1} \right) \\
\end{eqnarray*}

The on-site trial wave function is 

\begin{eqnarray}
&& |\Psi \rangle_i = \sum_{\mu \cdots \sigma} C_{\mu \cdots \sigma}
| \mu \cdots \sigma \rangle \nonumber \\
&& 
| \mu \cdots \sigma \rangle=
\prod_{\alpha = \mu, \ldots, \sigma}
{\rm Tr}[\Delta_\alpha \hat \psi^\dagger] |0 \rangle,
\nonumber \\
\end{eqnarray}
where $C_{\mu \cdots \sigma}$ is the amplitude at a particular state
$| \mu \dots \sigma \rangle$.
After tracing over the traceless tensors $\Delta_\mu$, we express the expectation
value of the Hamiltonian in Eq. (\ref{hamloc}) in terms of these amplitudes. 
By minimizing the energy with respect to $C_{\mu \cdots \sigma}$, we
obtain the ground states in different parameter regions and the mean-field
phase diagrams.

\section{Phase diagrams for nonzero tunneling}

In this section we present the results of numerical calculations following 
the scheme introduced 
in the previous section. We present result for two, three and four particles per 
lattice site. 

\subsection{$\rho=2$: Two particles per site}

\begin{figure}
\begin{center}
\includegraphics[scale=.66]{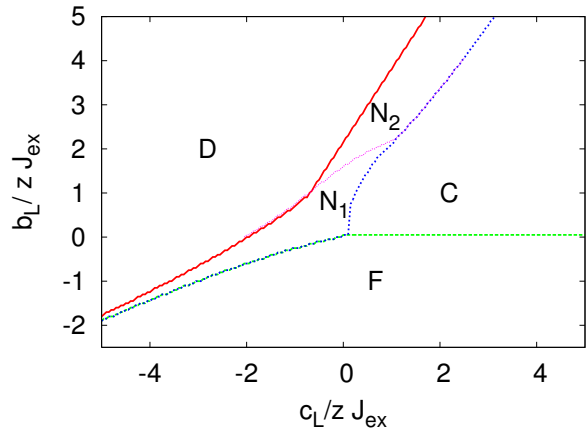}
\end{center}
\caption{(Color online) Phase diagram for two particles at a site: phase boundaries between the
dimer ($D$), nematic ($N_{1,2}$), cyclic ($C$) and ferromagnetic ($F$) phases}
\label{phasdiag2}
\end{figure}

For two particles per site we obtain the
phase diagram in Fig. \ref{phasdiag2}.
As pointed out before \cite{Zhou06}, in this case, dimer, nematic, cyclic and 
ferromagnetic phases appear. 
Moreover, we also 
observe an additional nematic phase between the dimer and cyclic phase;
in Fig. \ref{phasdiag2} this phase is indicated as $N_2$. This state differs from the $N_1$ state by its decomposition in terms of eigenstates of the total spin. The $N_1$ state has a nonzero projection in the $F=0$, $F=2$ and $F=4$ state, but the $N_2$ state consists only of states with $F=0$ and $F=2$. The $F=4$ components are absent because of the large value of $b_L$ at the position where this phase appears. We call the $N_1$ state a Maximally Ordered nematic State and the $N_2$ state a Minimally Ordered nematic State, because the $N_2$ state only involves the minimally needed states to break the translational symmetry.  

Quantifying the phase diagram we see that for $b_L$ and $c_L$ positive, 
but when $b_L/c_L < \tfrac{5}{3}$ the system remains in the cyclic phase. Upon increasing $b_L/c_L$ 
there is the possibility of a phase transition from the cyclic phase to the nematic phase as $J_{\rm ex}$ 
is varied and ultimately there is also a transition from the nematic phase to the dimer phase when $J_{\rm ex}$ is decreased. For small $J_{\rm ex}$ the phase boundary between the cylic and dimer phase approaches $b_L/c_L = 5/3$, in agreement with the analysis in Sec. III. We also find this agreement for the phase boundary between the ferromagnetic and dimer phase: it indeed approaches $b_L/c_L = 1/2$ for small $J_{\rm ex}$.

\begin{figure}
\begin{center} 
\includegraphics{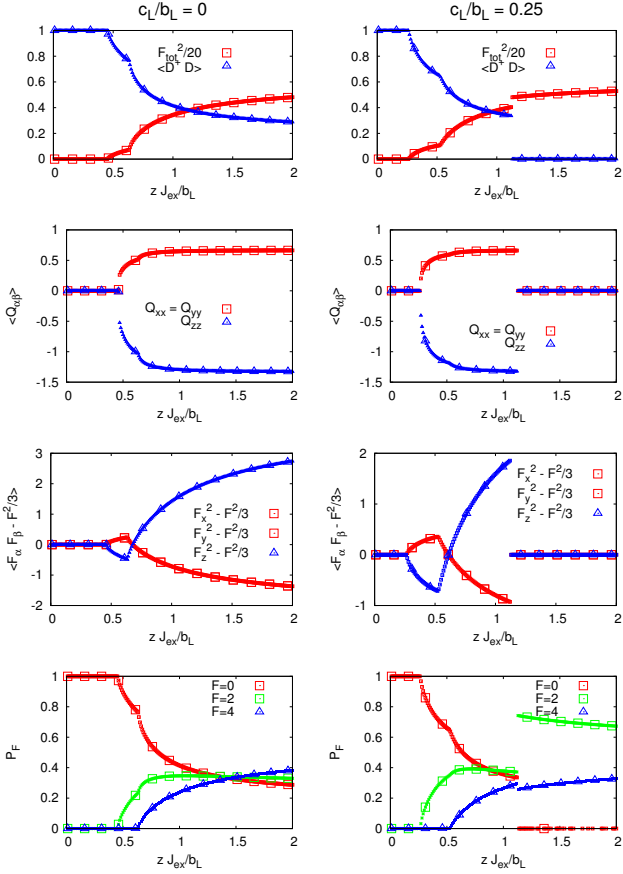}
\end{center}
\caption{(Color online) Order parameters for 2 particles per site and two ratio's of $c_L/b_L$. 
At the top row we plot the expectation value of the dimer counting operator $\hat{\mathcal{D}}^\dagger \hat{\mathcal{D}}$ and the total spin operator $\hat F^2$, which is rescaled with a factor $1/20$ for visual clarity. The second row displays the eigenvalues of the nematic order parameter $\hat Q_{\alpha \beta}$. Two eigenvalues ($Q_{xx}, Q_{yy}$) are always identical. The third row shows the eigenvalues of the operator $\hat F_\alpha \hat F_\beta - \frac{1}{3} \delta_{\alpha \beta} \hat F^2$. Again, two eigenvalues are identical. At the bottom row we plot the projection of the wave function onto the states with total spin $F=0,2,4$. The total spin states $F=1,3$ are not allowed because of the bosonic symmetry.
For $c_L/b_L=0$ (left column) we observe the transition from the dimer phase into the nematic (first $N_2$, then $N_1$) phase. For $c_L/b_L=0.25$ (right column) the system has an additional transition to the cyclic phase. 
}
\label{m2c0}
\end{figure}

In Figs. \ref{m2c0} order parameters are plotted for two ratio's of $c_L/b_L$. 
In particular we choose $c_L=0$ as realized for $^{87}$Rb \cite{Widera06} and $c_L/b_L=0.25$, 
as realized for $^{23}$Na. 
As is visible there, the $N_2$ nematic phase experiences a second order transition to the dimer phase. 
By contrast, the transition between the $N_1$ nematic phase and the dimer phase as shown in Fig. \ref{m2c0} is of 
first order. Also the transition between the $N_2$ nematic phase and the cyclic phase is of first order.

\begin{figure}
\begin{center} 
\includegraphics[scale=.66]{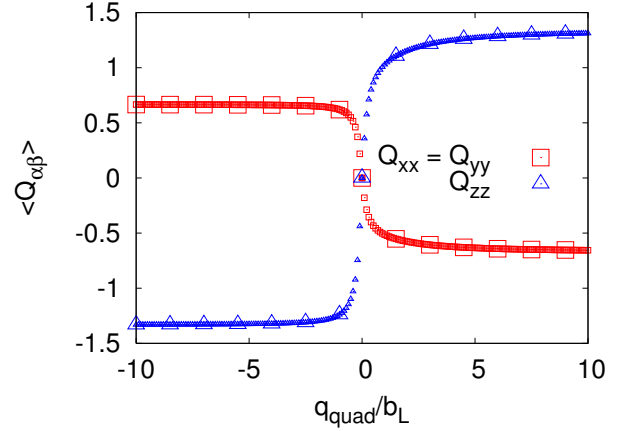}
\end{center}
\caption{(Color online) Eigenvalues of the nematic tensor in the presence of a quadratic 
Zeeman field for two particles per site in the dimer phase ($c_L/b_L = 0$, $z J_{\rm ex}/b_L = 0.2$) 
as a function of $q_{\rm quad}/b_L$. For nonzero $q_{\rm quad}$ the system always displays nematic order.}
\label{quadz2p}
\end{figure}

We now investigate the stability of the dimer phase in the presence of a quadratic Zeeman field. The result is presented in Fig. \ref{quadz2p}. As anticipated, the dimer phase is unstable towards a quadratic Zeeman field and nematic order is induced for infinitesimal couplings.

\subsection{$\rho=3$: Three particles per site}
\begin{figure}[b]
\begin{center}
\includegraphics[scale=.66]{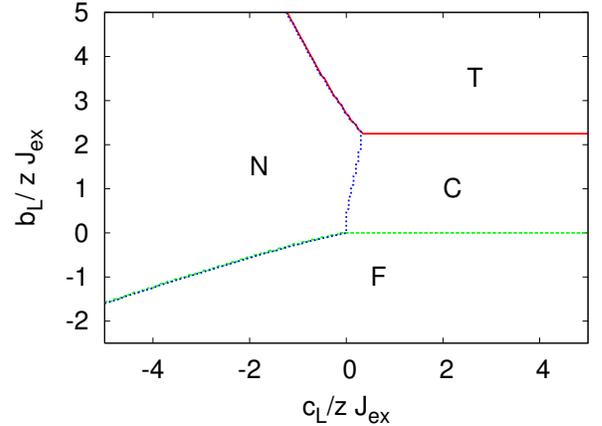}
\end{center}
\caption{(Color online) Phase diagram for three particles at a site: phase boundaries between the
trimer (T), nematic (N), cyclic (C) and ferromagnetic (F) phases}
\label{phasdiag3}
\end{figure}

For three particles per site we obtain the
phase diagram in Fig. \ref{phasdiag3}. As predicted before \cite{Zhou06} the phase diagram contains the nematic, cyclic, ferromagnetic and trimer phase. Calculating the phase border numerically, we see that the nematic phase extends into the positive $c_L$ quarter. In the asymptotic limit of small $J_{\rm ex}$ we recover the results from Sec. III that the critical slope separating the trimer and nematic phase is given by $b_L/c_L = - 7/3$ and between the nematic and ferromagnetic phase by $b_L/c_L = 7/18$.

\begin{figure}
\begin{center} 
\includegraphics{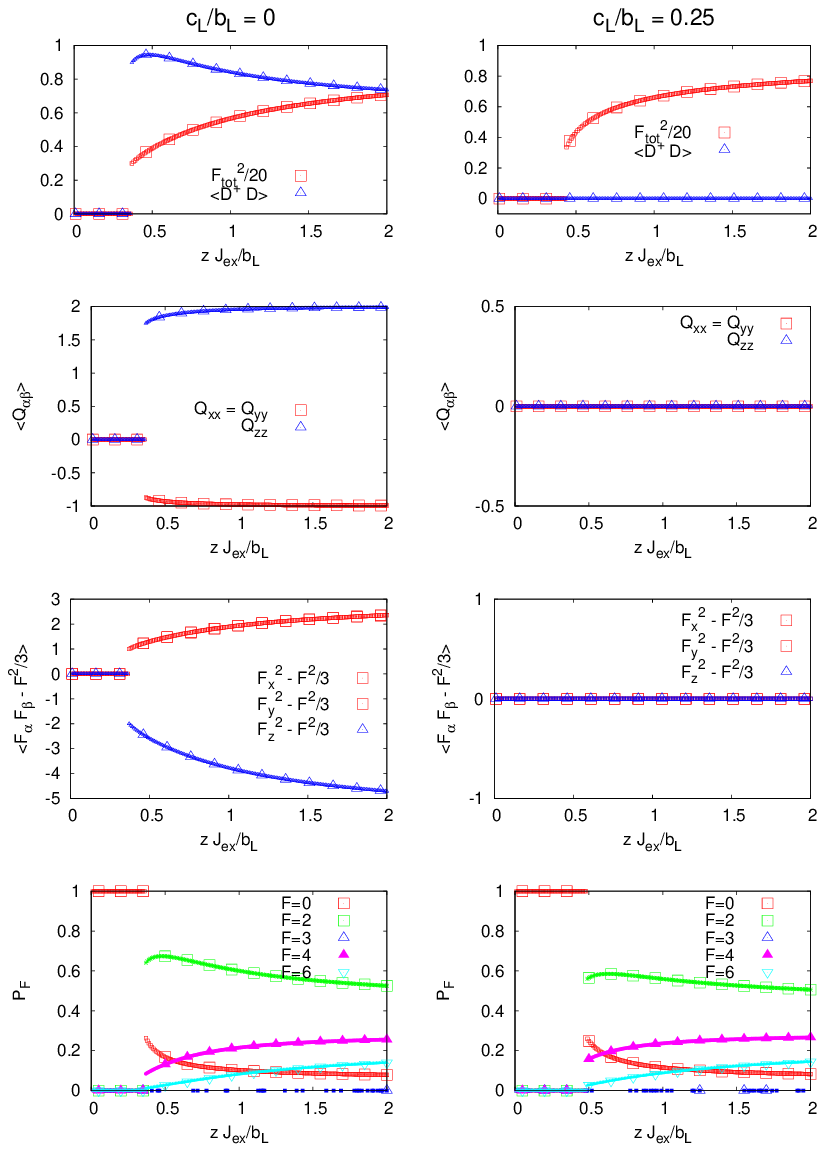}
\end{center}
\caption{(Color online) Order parameters for 3 particles per site and two ratio's of $c_L/b_L$. 
From top to bottom we plot expectation values ofthe dimer counting operator $\hat{\mathcal{D}}^\dagger \hat{\mathcal{D}}$ and the total spin operator $\hat F^2$, (top row); 
eigenvalues of the nematic order parameter $\hat Q_{\alpha \beta}$ (second row); 
eigenvalues of the operator 
$\hat F_\alpha \hat
F_\beta - \frac{1}{3} \delta_{\alpha \beta} \hat F^2$ (third row); and the projection of the wave function onto the states with total spin $F=0,2,3,4,6$ (bottom row). The total spin states $F=1,5$ are not allowed because of the bosonic symmetry.
For $c_L/b_L=0$ (left column) we observe the transition from the trimer phase into the nematic phase. For $c_L/b_L=0.25$ (right column) the system undergoes the trimer-cyclic transition. 
}
\label{m3c0}
\end{figure}

Again we present the order parameters for two ratio's of $c_L/b_L$ in Fig. \ref{m3c0}. From this we read off that the cyclic-trimer and nematic-trimer transition are both of first order nature. 

\begin{figure}
\begin{center} 
\includegraphics[scale=.66]{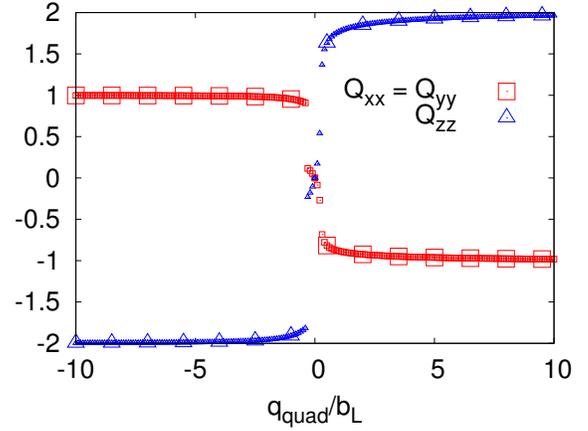}
\end{center}
\caption{(Color online) Eigenvalues of the nematic tensor in the presence of a quadratic Zeeman field for three particles per site in the trimer phase ($c_L/b_L = 0$, $z J_{\rm ex}/b_L = 0.2$) as a function of $q_{\rm quad}/b_L$. For nonzero $q_{\rm quad}$ there is always nematic order.}
\label{quadz3p}
\end{figure}

We also investigate the stability of the trimer phases against a quadratic Zeeman field. As shown in Fig. \ref{quadz3p}, the trimer phase is unstable against such a field, even for small values.

\subsection{$\rho=4$: Four particles per site}
\begin{figure}[b]
\begin{center}
\includegraphics[scale=.66]{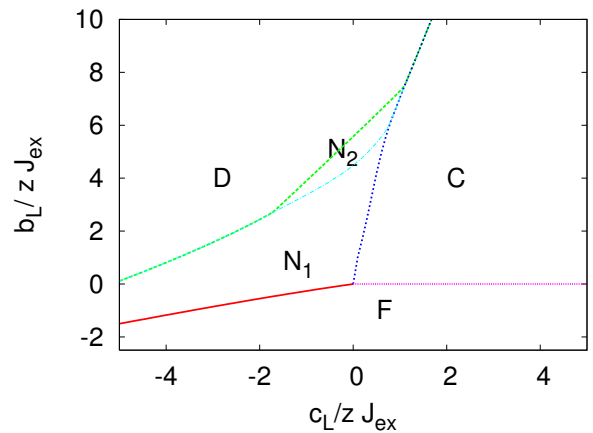}
\end{center}
\caption{(Color online) Phase diagram for four particles at a site: phase boundaries between the
dimer ($D$), nematic ($N_{1,2}$), cyclic ($C$) and ferromagnetic ($F$) phases}
\label{phasdiag4}
\end{figure}

For four particles per site we 
get the
phase diagram in Fig. \ref{phasdiag4}. Like in the case of two particles per site we obtain the ferromagnetic, cyclic, dimer and nematic phase.

\begin{figure}
\begin{center} 
\includegraphics{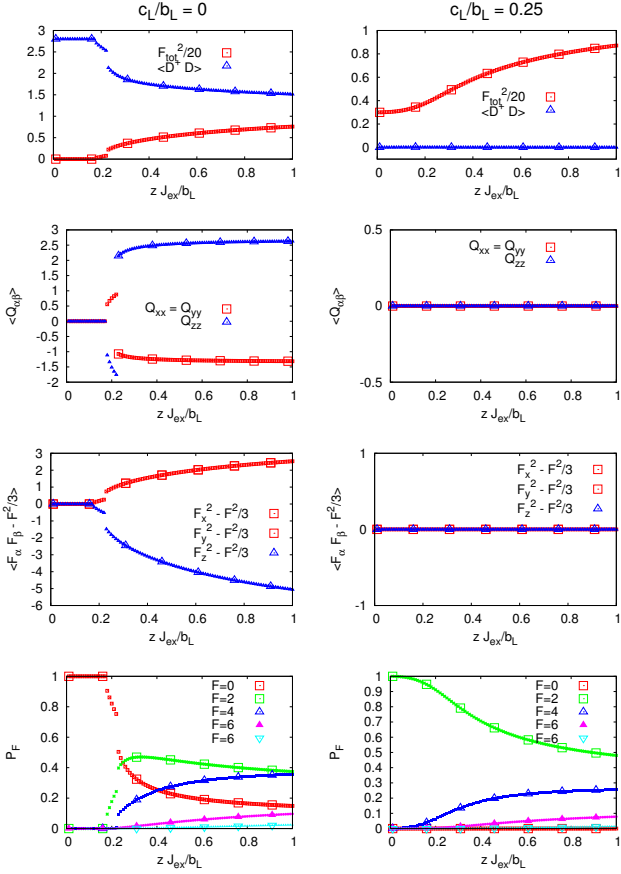}
\end{center}
\caption{(Color online) Order parameters for four particles per site and two ratio's of $c_L/b_L$. 
From top to bottom we plot expectation values of the dimer counting operator $\hat{\mathcal{D}}^\dagger \hat{\mathcal{D}}$ and the total spin operator $\hat F^2$ (top row); eigenvalues of the nematic order parameter $\hat Q_{\alpha \beta}$ (second row); eigenvalues of the operator $\hat F_\alpha \hat F_\beta - \frac{1}{3} \delta_{\alpha \beta} \hat F^2$ (third row); and the projection of the wave function onto the states with total spin $F=0,2,4,6,8$ (bottom row). The total spin states $F=1,3,7$ are not allowed because of the bosonic symmetry. Although $F=5$ is allowed, the states do not have a projection into this total spin state.
For $c_L/b_L=0$ (left column) we observe the transition from the dimer phase into the nematic (first $N_2$, then $N_1$) phase. For $c_L/b_L=0.25$ (right column) the system is always in the cyclic phase. 
}
\label{m4c0}
\end{figure}

Also in this case, the nematic phase is split into two sub-phases. The $N_2$ phase has only a projection into the 
$F=0$ and 
$F=2$ states, but the $N_1$ state has a projection into all the allowed total spin eigenstates. In analogy to the case for $\rho=2$, we call the $N_1$ a Maximally Ordered nematic State and the $N_2$ a Minimally Ordered nematic State. 
However, in contrast with the case of two particles per site, the nematic phase ($N_2$) only spreads over a finite 
area of the parameter space. For large 
$b_L/zJ_{\rm ex}$ there is a direct transition between the dimer phase and the cyclic phase.  For small $J_{\rm ex}$ the slope of the phase border between the dimer and cyclic phase is given by $b_L/c_L = 14/3$, in agreement with the analysis in Sec. III. Likewise the phase border between the ferromagnetic and dimer phase is given by $b_L/c_L = 7/18$ for small $J_{\rm ex}$. We present the order parameters for various ratio's of $c_L/b_L$ in Fig. \ref{m4c0}. It is clear that this gives a first order transition between the $N_1$ phase and the dimer phase. However, in this case also the transition between the $N_2$ phase and the 
dimer phase appears to be of first order.

\begin{figure}
\begin{center} 
\includegraphics[scale=.66]{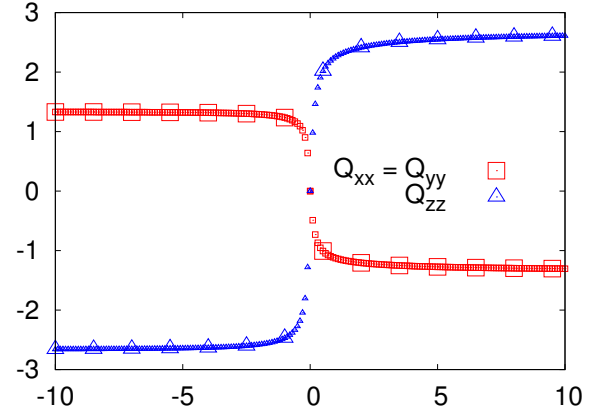}
\end{center}
\caption{(Color online) Eigenvalues of the nematic tensor in the presence of a quadratic Zeeman field for four particles per site in the dimer phase ($c_L/b_L = 0$, $z J_{\rm ex}/b_L = 0.1$) as a function of $q_{\rm quad}/b_L$. For nonzero $q_{\rm quad}$ there is always nematic order. 
}
\label{quadz4p}
\end{figure}

The stability of the dimer phase in a Quadratic Zeeman field is presented in Fig. \ref{quadz4p}. The dimer phase is unstable against a quadratic Zeeman field for infinitesimal fields.

\subsection{The case for the $F=2$ state of $^{87}$Rb}

We now turn to the experimentally most relevant case of $^{87}$Rb. For this case the parameters are such that $c_L = 0$ within experimental accuracy, whereas $a_L/b_L \approx 95$ \cite{Widera06}.
It is a particularly important question whether the dimer-nematic and trimer-nematic transitions happen within the Mott regime, i.e. whether on increasing the tunneling amplitude $t$ the transition from the dimer/trimer state occurs before the Mott insulating state is destroyed and the system becomes a superfluid with nematic order.  

In order to answer this question we calculate the
critical ratio $\frac{b_L}{z J_{\rm ex}}$ for which the spin-ordered to 
spin-disordered transition happens.
We then estimate the corresponding value of $a_L/t$ and compare it with  
the critical ratio $a_L/t$ at which the Mott insulator to superfluid 
transition occurs. We assume a three-dimensional cubic lattice and hence take $z=6$.
 
For two particles per site the Mott insulator to superfluid transition for spinless bosons occurs 
for $a_L/t \approx 100$ (note that $t$ as defined in this paper is half as large as normally used for spinless bosons) \cite{Santos09}. This has to be compared to the value of $a_L/t$ at the dimer-nematic 
transition, which happens when $\frac{b_L}{z J_{\rm ex}} \approx 2$ for $\rho = 2$. 
Taking into account  $J_{\rm ex} = \frac{t^2}{a_L}$,  
we conclude that the dimer nematic transition happens at 
$a_L/t \approx 34$ 
i.e. at a higher value of the hopping amplitude $t$ than 
the Mott-insulator superfluid transition. 
This means that for two particles per site the dimer-nematic transition in the Mott phase is preempted by the transition to the superfluid. 

For three particles per site the Mott insulator to superfluid transition for spinless bosons 
occurs for $a_L/t \approx 140$ \cite{Santos09}. This has to be compared to the trimer-nematic 
transition, which happens for $\frac{b_L}{z J_{\rm ex}} \approx 2.7$ for $\rho = 3$. This corresponds to 
$\frac{a_L}{t} \approx 39$. So the trimer-nematic transition won't take place before Mott states enter 
the superfluid phase

The superfluid-insulator transition for  $\rho=4$ happens at $\frac{a_L}{t} \approx 180$ \cite{Santos09}. As seen from Fig. \ref{m3c0} the dimer-nematic transition occurs for $\frac{b_L}{z J_{\rm ex}} \approx 5$, which corresponds to 
$
\frac{ a_L}{t} \approx  53 
$. This means that also for four particles per site the dimer-nematic transition is preempted by the Mott- insulator superfluid transition.

However, since the spin-ordering affects the phase boundary to the superfluid phase \cite{Hou03, Jin04}, 
the precise nature of these transtitions remains unclear and further investigation is needed.




\section{Conclusions}
In this article we studied magnetic transitions in the Mott states of spin-2 atoms for various particle numbers per site. We derived the exact phase diagram for zero tunneling. For the case of nonzero tunneling we used a self-consistent mean-field technique to study the phase diagram. We found various symmetry breaking transitions, depending on the microscopic parameters. In particular, for the microscopic parameters of $^{87}$Rb there is the possibility of a dimer-nematic transition and also a transition within the nematic phase, which corresponds to a transition between a maximally ordered nematic state and a minimally ordered nematic state. However, the nematic-dimer transition happens already for smaller $a_L/z J_{\rm ex}$ ratio than the usual Mott transition does. Therefore the precise nature of this transition remains to be clarified in future work.

A magnetic field induces a quadratic Zeeman coupling, which within the states with unbroken symmetry gives rise to nematic order even for infinitesimal coupling.  

\section*{Acknowledgements}
This work is supported by the Nederlandse Organisatie voor Wetenschappelijk Onderzoek (NWO) and by NSERC, the Canada and Canadian Institute for Advanced Research

\end{document}